\documentclass[usenatbib]{mn2e}

\usepackage{dcolumn}
\usepackage{graphics,epsfig}
\usepackage{txfonts}
\usepackage{times}
\usepackage{pslatex}
\usepackage{amssymb}

\def\hi{H{\sc i}}

\newcommand{\cm}{cm$^{-2}$}

\newcommand{\ts}{T_s}

\newcommand{\beq}{\begin{equation}}
\newcommand{\eeq}{\end{equation}}

\title[The covering factor of high redshift damped Lyman-$\alpha$ systems]{The covering factor of high redshift damped Lyman-$\alpha$ systems}
\author[Kanekar et al.]{N.~Kanekar$^1$\thanks{E-mail: nkanekar@nrao.edu (NK)},
W.~M.~Lane$^2$, E.~Momjian$^1$, F.~H.~Briggs$^{3}$, J.~N.~Chengalur$^{4}$\\
$^1${}National Radio Astronomy Observatory, 1003 Lopezville Rd, Socorro, NM 87801, USA; \\
$^2${}Naval Research Laboratory, Code 7213, 4555 Overlook Ave SW, Washington, DC 20375, USA\\
$^3${}Australian National University, ACT 2611, Australia \\
$^4${}National Centre for Radio Astrophysics, Ganeshkhind, Pune--411007, India}

\begin{document}
\date{Received mmddyy/ accepted mmddyy}
\maketitle
\label{firstpage}

\begin{abstract}
We have used the Very Long Baseline Array to image 18~quasars with foreground damped Lyman-$\alpha$ 
systems (DLAs) at 327, 610 or 1420~MHz, to measure the covering factor $f$ of each DLA at or 
near its redshifted \hi~21cm line frequency. Including six systems from the literature, we 
find that none of 24~DLAs at $0.09 < z < 3.45$ has an exceptionally low covering factor, 
with $f \sim 0.45 - 1$ for the 14~DLAs at $z > 1.5$, $f \sim 0.41 - 1$ for the 10~systems at $z < 1$, 
and consistent covering factor distributions in the two sub-samples. The observed 
paucity of detections of \hi~21cm absorption in high-$z$ DLAs thus cannot 
be explained by low covering factors and is instead likely to arise due to a larger 
fraction of warm \hi\ in these absorbers.  
\end{abstract}


\section{Introduction}
\label{intro}

Damped Lyman-$\alpha$ systems (DLAs), selected on the basis of their high \hi\ column densities 
($N_{\rm HI} \ge 2 \times 10^{20}$~\cm) in quasar absorption spectra, have long been identified 
as the progenitors of normal present-day galaxies \citep{wolfe86}, and have hence been the 
subject of much research. Despite this, the nature of high-$z$ DLAs, physical conditions 
in them, and their evolution with redshift, are still matters of dispute today (e.g. 
\citealp{wolfe05}).  An issue of recent controversy is the temperature distribution of \hi\ in DLAs. 
For radio-loud background quasars, a comparison between the \hi\ column density and 
the optical depth in the redshifted \hi~21cm line yields the DLA spin temperature ($\ts$). 
For optically-thin multi-phase absorption, $\ts$ is the column-density-weighted harmonic 
mean of the spin temperatures of the different phases along the line of sight.
For nearly three decades, the spin temperatures of high-$z$ DLAs have been found to be
systematically higher than values seen in the Milky Way or local spiral disks (e.g. 
\citealp{wolfe79,carilli96,kanekar03}); the first low-$\ts$ DLA at high redshift was
only detected very recently \citep{york07}. The simplest interpretation of the observations 
is that \hi\ in high-$z$ DLAs is predominantly warm, unlike the situation in the Galaxy 
\citep{carilli96,chengalur00,kanekar03}. Conversely, \citet{wolfe03b} used C{\sc II}* 
absorption lines to argue that roughly half the DLAs at $z \gtrsim 2$ contain significant 
fractions of cold \hi.

A problem in measuring DLA spin temperatures is that low-frequency radio emission 
is often extended over large angular scales, far larger than the size of
a galaxy. Redshifted \hi~21cm absorption studies are usually carried out with single 
dishes or short-baseline interferometers, of fairly poor angular resolution, 
$\gtrsim 10''$, or $\gtrsim 80$~kpc at $z \sim 3$. It is hence often not certain that 
the quasar radio emission is entirely covered by the foreground 
DLA. If some fraction of the quasar emission leaks out around the DLA, the inferred spin temperature 
will be an over-estimate. The fraction of the radio emission occulted by the DLA is 
referred to as the covering factor $f$ (e.g. \citealp{briggs83}). 

Recently, \citet{curran05} have emphasized the problem of unknown covering factors in 
high-$z$ DLAs, pointing out that the high estimated $\ts$ values of \citet{kanekar03} (hereafter KC03)
could merely stem from very low DLA covering factors, $f << 1$.  \citet{curran06} also 
argued that low DLA covering factors at high redshifts arise due to a geometrical effect, 
as the similarity in angular diameter distances of the high-$z$ DLAs and their background quasars
implies that a high-$z$ DLA is ``less effective'' at covering the quasar than a low-$z$ absorber. 
Note that \citet{curran05} find that quasars of angular extent $\lesssim 0.1''$ are adequately 
covered even by compact foreground DLAs.

The most direct way of resolving this issue is by high spatial resolution very long baseline 
interferometry (VLBI) studies in the redshifted \hi~21cm line (e.g. \citealp{briggs89,lane00}). 
Unfortunately, the low-frequency coverage of current VLBI facilities is quite limited, implying 
that such observations are only possible for very few sources. Alternatively, one could
measure the compact flux density arising from the quasar core in VLBI continuum 
images at or near the redshifted \hi~21cm line frequency and then estimate the DLA covering 
factor by comparing the VLBI flux density with the flux density measured with short baseline 
interferometers or single dishes (e.g. \citealp{briggs83,kanekar07}).  This is the approach
that we follow here, based on Very Long Baseline Array (VLBA) imaging of 18~quasars 
with foreground DLAs at $z \sim 0.24 - 3.45$.

\section{Observations, data analysis and results}
\label{sec:obs}

\setcounter{figure}{0}
\begin{figure*}
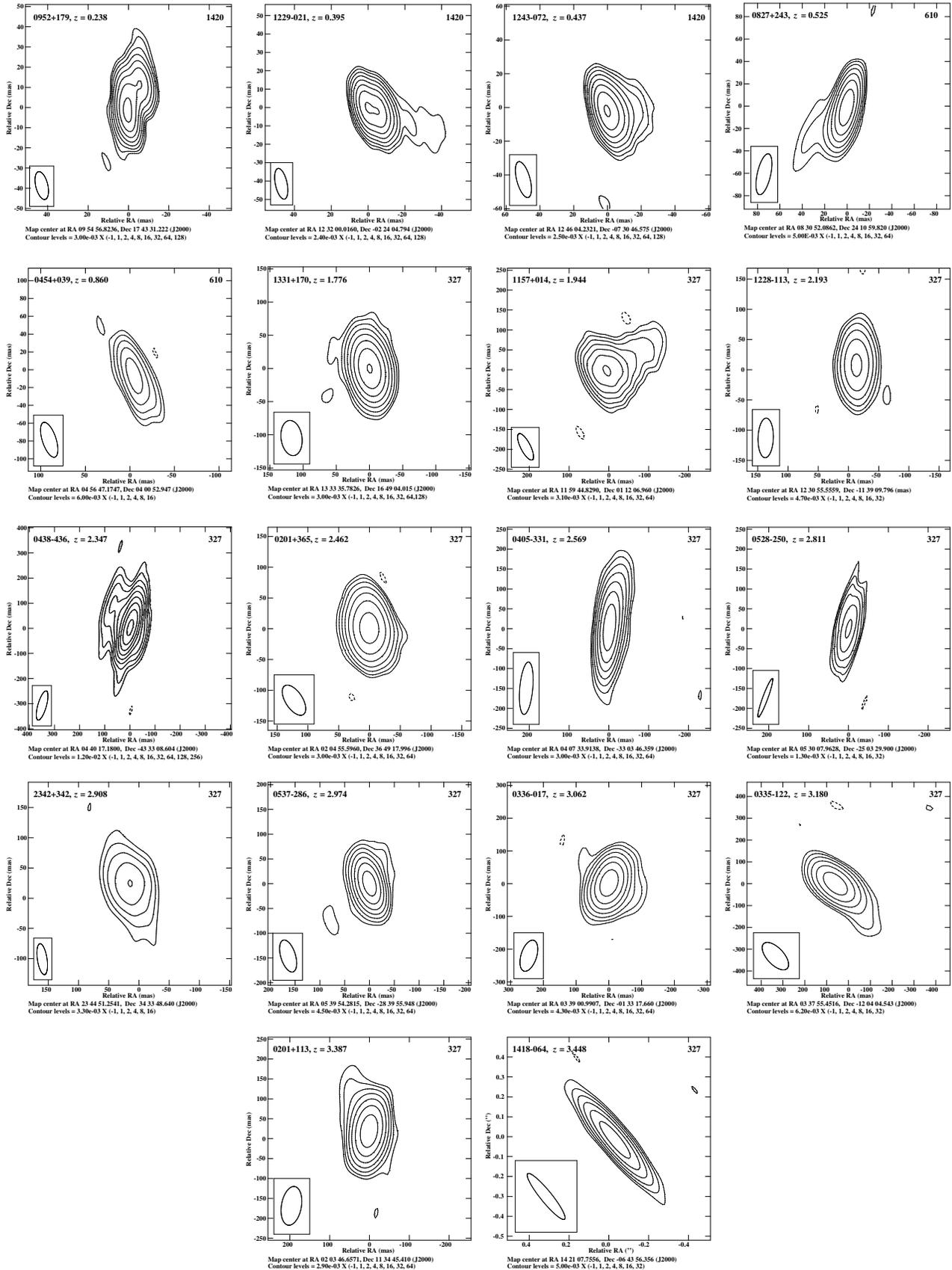

\begin{center}
\epsfig{file=fig1a.eps,width=1.65in}
\epsfig{file=fig1b.eps,width=1.65in}
\epsfig{file=fig1c.eps,width=1.65in}
\epsfig{file=fig1d.eps,width=1.65in}
\epsfig{file=fig1e.eps,width=1.65in}
\epsfig{file=fig1f.eps,width=1.65in}
\epsfig{file=fig1g.eps,width=1.65in}
\epsfig{file=fig1h.eps,width=1.65in}
\epsfig{file=fig1i.eps,width=1.65in}
\epsfig{file=fig1j.eps,width=1.65in}
\epsfig{file=fig1k.eps,width=1.65in}
\epsfig{file=fig1l.eps,width=1.65in}
\epsfig{file=fig1m.eps,width=1.65in}
\epsfig{file=fig1n.eps,width=1.65in}
\epsfig{file=fig1o.eps,width=1.65in}
\epsfig{file=fig1p.eps,width=1.65in}
\epsfig{file=fig1q.eps,width=1.65in}
\epsfig{file=fig1r.eps,width=1.65in}
\end{center}
\vskip -0.1in
\caption{VLBA images of the 18~quasars of the sample, ordered (from top left) by DLA 
redshift. The map frequency is at the top right of each panel.}
\label{fig:images}
\end{figure*}

The VLBA 1.4~GHz, 610~MHz and 327~MHz receivers were used to observe the 18~quasars of
our sample between 2002~March and 2006~June (projects BK89 and BK131), with observing 
frequencies close to the redshifted \hi~21cm line frequency of the foreground DLA. 
Three quasars with DLAs at $z \sim 0.24 - 0.44$ were observed at 1.4~GHz, two, with DLAs at 
$z \sim 0.52 - 0.86$, at 610~MHz, and 13, with DLAs at $z \sim 1.78 - 3.45$, at 327~MHz. 
Total bandwidths of 4, 8 and 16~MHz were used for the observations at 610~MHz, 327~MHz and 
1420~MHz, respectively, with 2-bit sampling, two polarizations, 16~spectral channels and on-source 
integration times of $1 - 4$~hours.  Strong fringe finders (3C345, 3C454.3 or 3C286) were 
observed every few hours, to calibrate the shape of the passband. Phase-referencing was not used.
The U-V coverage was poor in some cases, as all the VLBA antennas were not available,
resulting in asymmetric or larger synthesized beams.

The data were analysed in ``classic'' {\sc AIPS}, using standard techniques.  Initial 
procedures included ionospheric corrections, editing of radio frequency interference, 
amplitude calibration (using the measured antenna gains and system temperatures), 
passband calibration and fringe-fitting for the delay rates, before the data for each 
target were averaged in frequency to a single-channel dataset. For each source, this was 
followed by a number of cycles of self-calibration and imaging to determine the antenna gains, 
until no improvement was seen on further self-calibration. The final images obtained from the above 
procedure are shown in Fig.~\ref{fig:images}, in order of increasing redshift, with the synthesized 
beams listed in column~(6) of Table~\ref{tab:fits}. The root-mean-square noise in off-source regions 
in the images was measured to be $\sim 0.5 - 3$~mJy/Bm. 

The task {\sc UVFIT} was used to fit elliptical gaussian models to the visibility data,
to measure the compact flux density. Most sources in Fig.~\ref{fig:images} are 
core-dominated, with only a small fraction of the flux density in extended structure.
For some sources with visible extensions (e.g. 1229$-$021, 1243$-$072 and 0201+113), a 
single-component model was found sufficient to recover the ``cleaned'' flux density. The only
exceptions are 0952+179~($z_{\rm abs} \sim 0.238$), 1157+014~($z_{\rm abs} \sim 1.944$) and 
0438$-$436~($z_{\rm abs} \sim 2.347$), where all the ``cleaned'' flux density could 
be recovered only with a 2-component model.

Table~\ref{tab:fits} summarizes the results of the VLBA observations, grouping the sources by observing 
band, in order of increasing redshift. For each source, the first four columns contain the observing band, 
source name, DLA redshift and redshifted \hi~21cm frequency. Column~(5) lists the total flux densities at 
the VLBA observing frequency. For the three 1.4~GHz targets, this was obtained from the 1.4~GHz NRAO 
VLA Sky Survey \citep{condon98}. For most 327~MHz targets (11/13), we use the flux density measured at 
nearby frequencies [e.g. at 327~MHz, from the Westerbork Northern Sky Survey \citep{rengelink97}, 365~MHz, from 
the Texas survey \citep{douglas96}, \citet{carilli96} or KC03]. Four sources (0827+243, 0454+039, 0405$-$331 
and 0438$-$436) do not have measurements in the literature near the VLBA observing frequencies; their 
flux densities were hence estimated from their low-frequency ($< 1$~GHz) spectral indices. 

Columns~($6-10$) contain (6)~the VLBA synthesized beam, (7)~the flux density obtained on fitting 
a gaussian model with {\sc UVFIT}, (8)~the deconvolved angular sizes of the gaussian components, 
(9)~the corresponding spatial extents of these components, at the redshift of the foreground DLA 
[we use $(\Omega_\Lambda, \Omega_m, h) = (0.7,0.3,0.7)$ in this paper] and (10)~the covering factor 
of the compact radio emission at the observing frequency, obtained by dividing the ``core'' flux 
density by the total flux density. In three cases of two source components, the one that is more compact 
is identified with the core. We emphasize that the deconvolved sizes listed in columns~(8) and 
(9) are upper limits, due to the possibility of residual phase errors in the data (e.g. due to 
fluctuations on time scales shorter than the self-calibration interval). This is especially 
true for the 327~MHz and 610~MHz results. 

\setcounter{table}{0}
\begin{table*}
\begin{center}
\begin{tabular}{|c|c|c|c|c|c|c|c|c|c|c|}
\hline
$\nu_{\rm obs}$& QSO & $z_{\rm abs}$ & $\nu_{\rm 21cm}$ & $S_{\rm tot}$ & Beam             & S$_{\rm fit}$ & Angular size & Spatial extent  & $f$\\
     &     &               & MHz   & Jy            & mas $\times$ mas &   mJy         &  mas $\times$ mas       & pc $\times$ pc         & \\
\hline
1.4 & 0952+179   & 0.238 & 1147.5 & 1.16 & $14 \times 6$ & $762 \pm 1$ & $(1.7 \pm 0.1) \times (13.9 \pm 0.1)$ & $(6.4 \pm 0.1) \times (52.2 \pm 0.1)$ & 0.66 \\
  &        &       &        &      &               & $303 \pm 1$ & $(4.5 \pm 0.1) \times (9.1 \pm 0.1)$ & $(17.1 \pm 0.1)\times (34.1 \pm 0.1)$ & $-$ \\ 
1.4 & 1229$-$021 & 0.395 & 1018.2 & 1.65 & $17 \times 6$ & $689 \pm 1$ & $(14.9 \pm 0.1) \times (2.5\pm 0.1)$ & $(79.6\pm 0.1) \times (13.1 \pm 0.8)$ & 0.42 \\
1.4 & 1243$-$072 & 0.437 & 988.6  & 0.55 & $22 \times 8$ & $483 \pm 1$ & $(7.7 \pm 0.1) \times (6.9 \pm 0.1)$ & $(43.6 \pm 0.2) \times (39 \pm 1)$ & 0.88 \\
&&       &        &      &                 &                 &                                 & 		&\\
610 & 0827+243   & 0.525 & 931.6  & 0.90 & $38 \times 13$& $626 \pm 6$ & $(2.3 \pm 0.4) \times (11.3 \pm 0.4)$ & $(16 \pm 2) \times (71 \pm 3)$ & 0.70 \\
610 & 0454+039   & 0.860 & 763.8  & 0.50 & $41 \times 15$  & $251 \pm 9$ & $(31 \pm 1) \times (9 \pm 1)$ & $ (241 \pm 9) \times (66 \pm 9)$ & 0.50 \\
&&       &        &      &                 &                 &                                 & 		&\\
327 & 1331+170   & 1.776 & 511.6  & 0.62 & $53 \times 32$  & $444 \pm 5$ & $(26 \pm 1) \times (2 \pm 2)$ & $(223 \pm 3) \times (19 \pm 14)$  & 0.72 \\
327 & 1157+014   & 1.944 & 482.5  & 0.89 & $74 \times 25$  & $565 \pm 15$    & $ (39 \pm 1) \times (58 \pm 1)$ & $(327 \pm 8)\times (488 \pm 9)$ & 0.63 \\
  &        &       &        &      &                 & $132 \pm 13$    & $ (3 \pm 35) \times (107\pm 8)$ & $(23 \pm 297)\times (897 \pm 64)$ & $-$ \\ 
327 & 1228$-$113 & 2.193 & 444.8  & 0.51 & $65 \times 25$  & $290 \pm 9$ & $(39 \pm 2) \times (24 \pm 1)$ & $(322 \pm 18)\times (196 \pm 6)$ & 0.57 \\
327 & 0438$-$436 & 2.347 & 424.4  & 7.64 & $120 \times 34$ & $4533 \pm 15$   & $(19 \pm 1) \times (34 \pm 1)$ & $(156 \pm 2) \times (277 \pm 2)$ & 0.59 \\
  &        &       &        &      &                 & $1038 \pm 18$   & $(123 \pm 2) \times (298 \pm 12)$  & $(1008 \pm 11)\times (2434 \pm 101)$ &  $-$ \\
327 & 0201+365   & 2.462 & 410.3  & 0.52 & $56 \times 28$  & $510 \pm 6$ & $(10 \pm 1) \times (51.1 \pm 0.4)$ & $(79 \pm 4) \times (414 \pm 3)$ & 0.98 \\
327 & 0405$-$331 & 2.569 & 398.0  & 0.83 & $132 \times 33$ & $367 \pm 5$ & $(7 \pm 1) \times (0 \pm 2) $ & $(55 \pm 6)\times (0 \pm 15)$ & 0.44 \\
327 & 0528$-$250 & 2.811 & 372.7  & 0.14 & $102 \times 13$ & $132 \pm 5$ & $(49 \pm 5) \times (8 \pm 3))$ & $(384 \pm 42) \times (65 \pm 20)$ & 0.94\\
327 & 2342+342   & 2.908 & 363.5  & 0.31 & $52 \times 16$  & $219 \pm 10$ & $(49 \pm 2) \times (42 \pm 2)$& $(372 \pm 13) \times (329 \pm 12)$ & 0.71 \\
327 & 0537$-$286 & 2.974 & 357.4  & 1.05 & $67 \times 30$  & $495 \pm 4$ & $(10 \pm 1) \times (10 \pm 1) $ & $(77 \pm 2)\times (79 \pm 10)$  & 0.47 \\
327 & 0336$-$014 & 3.062 & 349.7  & 0.94 & $99 \times 48$  & $636 \pm 7$ & $(37 \pm 1) \times (23 \pm 2)$ & $(286 \pm 2) \times (176 \pm 18)$ & 0.68 \\
327 & 0335$-$122 & 3.180 & 339.8  & 0.68 & $157 \times 77$ & $419 \pm 8$ & $(31 \pm 6) \times (69 \pm 2) $& $(233 \pm 45) \times (521 \pm 11)$ & 0.62 \\
327 & 0201+113   & 3.387 & 323.8  & 0.42 & $98 \times 50$  & $321 \pm 8$ & $(21 \pm 2) \times (6 \pm 4)$ & $ (152 \pm 12) \times (41 \pm 28)$ & 0.76 \\
327 & 1418$-$064 & 3.448 & 319.3  & 0.44 & $294 \times 57$ & $302 \pm 11$& $(23 \pm 33) \times (38 \pm 2) $& $(167 \pm 240) \times (280 \pm 11)$ & 0.69\\
\hline
\end{tabular}
\caption{Results from VLBA low-frequency imaging of quasars behind high-$z$ DLAs. See text for details and discussion.}
\label{tab:fits}
\end{center}
\vskip -0.1in
\end{table*}

\section{Other sources}
\label{sec:others}

A few other quasars with foreground DLAs have estimates of the covering factor from VLBI studies.
These are discussed below.\\
(1)~0738+313 ($z_{\rm abs} \sim 0.0912$, $0.2212$): \citet{lane00} used 1302~MHz VLBA observations
to estimate that the covering factor of the foreground DLAs is $f \sim 0.98$.\\
(2)~1127$-$145 ($z_{\rm abs} \sim 0.3127$): \citet{bondi96} measured a flux density of $\sim 5.0$~Jy in their 
1.6~GHz VLBI image, compared to a single-dish flux density of $\sim 5.6$~Jy. This gives $f \sim 0.89$.\\
(3)~0235+164 ($z_{\rm abs} \sim 0.524$): \citet{wolfe78} used 931~MHz VLBI observations to show that the 
size of the quasar core is $< 6$~mas.
More recently, \citet{frey00} found that the 1.6~GHz VLBI core flux density (with a sub-mas beam) is 
very similar to that measured simultaneously with a single dish. This implies $f \sim 1$.\\
(4)~3C286 ($z_{\rm abs} \sim 0.692$): \citet{wilkinson79} used 609~MHz VLBI observations to find that 17.5~Jy 
of the 609~MHz flux density ($\sim 20$~Jy) arises in the central $55$~mas. This implies $f \gtrsim 0.9$.\\
(5)~0458-020 ($z_{\rm abs} \sim 2.039$): \citet{briggs89} used 608~MHz VLBI observations to estimate that the compact
quasar core contains $\sim 1.15$~Jy at the redshifted \hi~21cm line frequency of $\sim 467$~MHz, with an additional
$\sim 1.3$~Jy in two extended components, on scales of $\sim 0''.2 - 0''.5$ and $\sim 1-2''$. 
They found the \hi~21cm optical depths measured in VLBI and single-dish studies to be in excellent 
agreement and used this to argue that the entire $2''$ radio emission is likely to be covered, i.e. 
$f \sim 1$.\\
There are four DLAs with searches for \hi~21cm absorption where the core fraction in the 
background quasar is very small, implying a low covering factor ($f \lesssim 0.1$). 
These are at $z \sim 0.437$ towards 3C196, $z \sim 1.3911$ towards 0957+561A, $z \sim 1.4205$ towards 
1354+258 and $z \sim 0.656$ towards 3C336. \citet{boisse98} noted that the optical and radio sightlines 
towards 3C196 are clearly different, due to which $\ts$ cannot be estimated. Similarly, 
KC03 found very small core fractions in 0957+561A and 1354+258, and argued that covering 
factor uncertainties preclude an estimate of the spin temperature. Conversely, 
\citet{curran07a} do quote a (low) spin temperature for the DLA towards 3C336. However, the quasar 
is strongly lobe-dominated with the lobes extended over $\sim 190$~kpc, and only a small fraction 
of the flux density arises from the core \citep{bridle94}. As in the case of 3C196, the difference 
between radio and optical sightlines implies that it is not possible to estimate $\ts$ in this absorber.

\section{Discussion}
\label{sec:discuss}

For \hi~21cm absorption studies of DLAs towards radio-loud quasars, the \hi~21cm 
optical depth $\tau_{\rm 21}$, \hi\ column density $N_{\rm HI}$ (\cm) and spin 
temperature $\ts$ (K) are related by the expression 
\begin{equation}
N_{\rm HI} = 1.823 \times 10^{18} \times \left( \ts/f \right) \int \tau_{\rm 21} {\rm d}V  \:\:,
\end{equation}
where $f$ is the covering factor of the foreground DLA and the \hi~21cm line is assumed to be
optically thin. For a DLA, where $N_{\rm HI}$ is known from the Lyman-$\alpha$ profile, a 
measurement of the \hi~21cm optical depth hence yields $\left(\ts/f\right)$. Estimates of the 
covering factor can then be used to infer $\ts$ for the DLA. High-$z$ DLAs have long been found 
to have higher $\ts$ values than seen in the Galaxy or local spirals (e.g. 
\citealp{wolfe79,carilli96}; KC03). For example, KC03 found that all seven DLAs in their high-$z$ 
sample had $\ts > 700$~K, while more than 80\% of sightlines through the Milky Way and M31 have 
$\ts < 350$~K \citep{braun92}. This suggests that high-$z$ DLAs have far smaller fractions of the 
cold phase of \hi\ than local spirals (see also \citealp{carilli96,kanekar01a}). However, the 
$\ts$ estimates depend on the estimated covering factors. If, as argued by \citet{curran05} and 
\citet{curran06}, high-$z$ DLAs have systematically low covering factors, the high $\left(\ts/f\right)$ 
measurements could arise due to low $f$ values, and not due to high spin temperatures. Note that 
5/7 high-$z$ DLAs of the KC03 sample have $\left[\ts/f\right] \gtrsim 2000$~K; extremely low 
covering factors ($f < 0.2$) would then be needed to obtain spin temperatures in the ``low'' Galactic 
range ($\ts < 350$~K).

The covering factor of a DLA can be determined from VLBI studies at the redshifted \hi~21cm line frequency. 
Such studies are fairly common at frequencies $\gtrsim 1.4$~GHz, implying that it is usually straightforward 
to determine $\ts$ for low-$z$ DLAs ($z \lesssim 0.5$), for which the \hi~21cm line is redshifted to 
frequencies $\gtrsim 1$~GHz. However, for DLAs at $z \gtrsim 1.7$, the line frequency
is redshifted to $\lesssim 500$~MHz, where it is technically difficult to carry out VLBI studies, as 
the coherent integration times are restricted by ionospheric fluctuations \citep{briggs83b}. This has meant 
that most $\ts$ measurements in high-$z$ DLAs are based on covering factors estimated either from VLBI studies 
at much higher frequencies (typically $\gtrsim 1.4$~GHz) or from the spectral index of the background quasar (e.g.  
\citealp{wolfe79,wolfe81,bruyn96,carilli96,kanekar03,kanekar06}). Prior to this work, there were only 
three DLAs at $z \gtrsim 1.7$ with VLBI studies at $\lesssim 600$~MHz
\citep{briggs83,briggs89}, only one of which used more than two VLBI stations.  

\begin{figure*}
\begin{center}
\epsfig{file=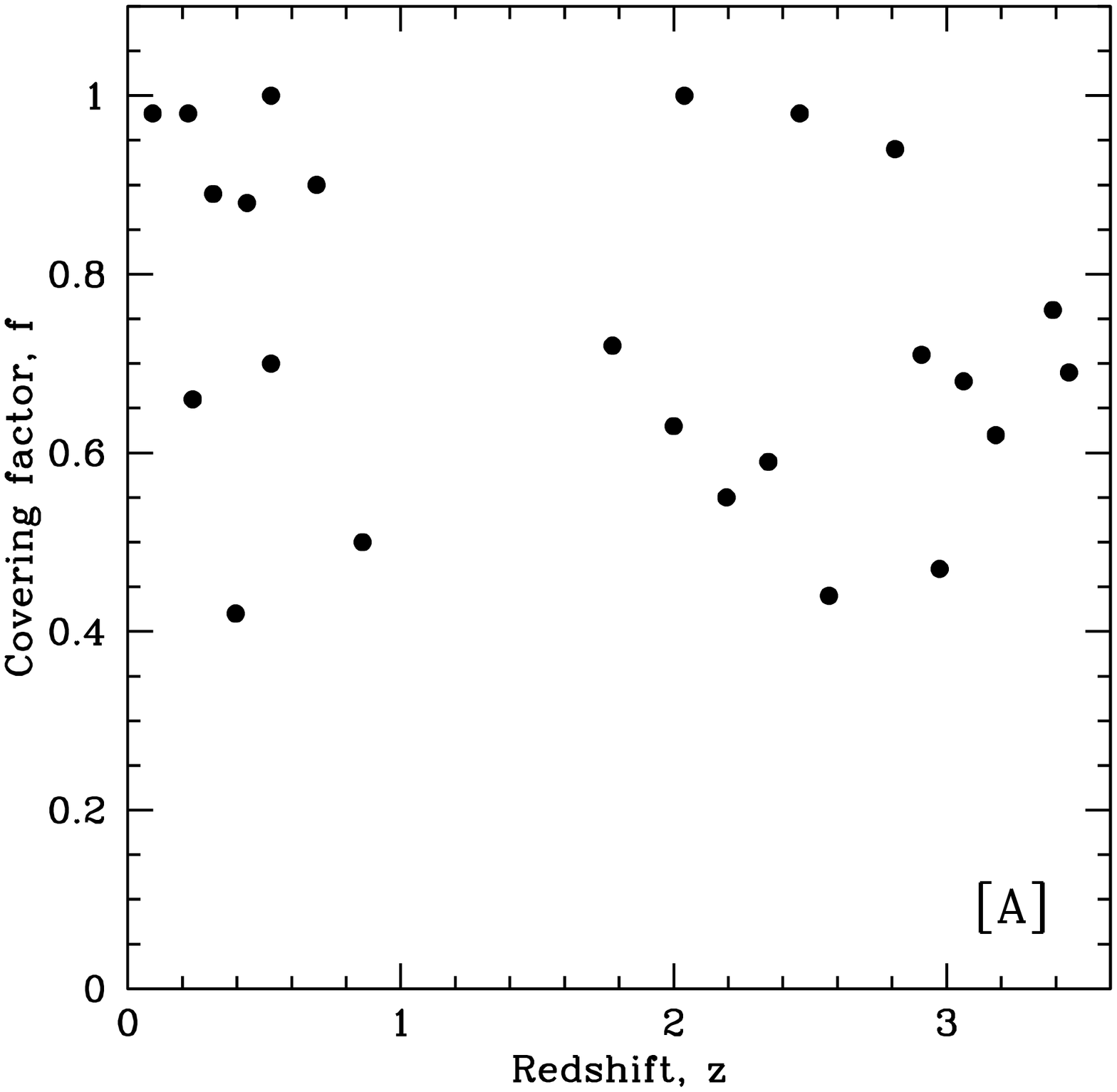,width=3.3in}
\epsfig{file=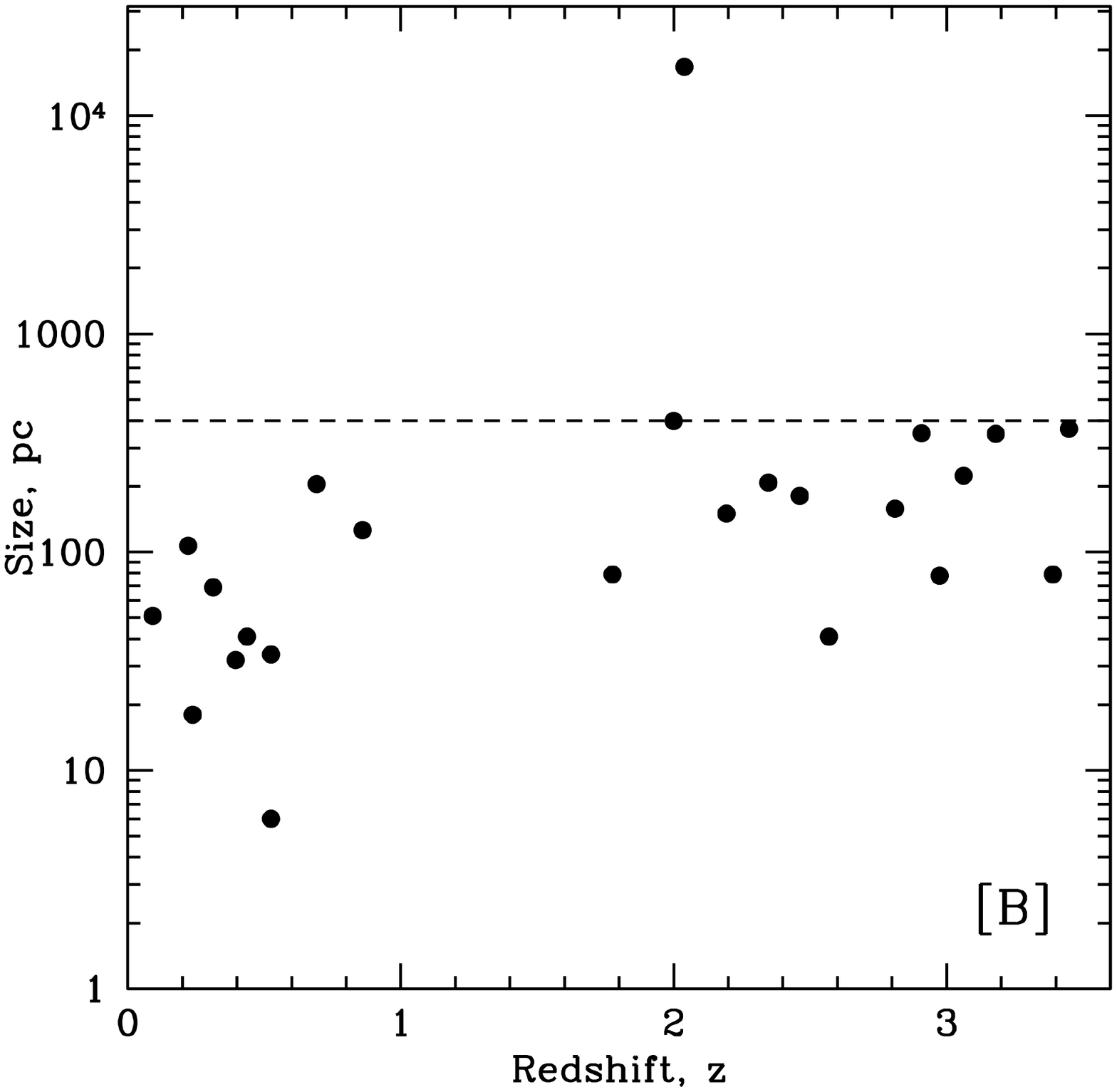,width=3.3in}
\end{center}
\vskip -0.15in
\caption{[A]~Left panel: The covering factor of the 24 DLAs of the present VLBI sample, plotted against
DLA redshift, $z_{\rm abs}$. [B]~Right panel: The spatial extent of the compact radio emission of the background 
quasars (at $z_{\rm abs}$) plotted versus $z_{\rm abs}$. The dashed line in [B] is at 400~pc. See text for 
discussion.}
\label{fig:f}
\vskip -0.1in
\end{figure*}

The full sample of quasars with DLA covering factor estimates from VLBI observations now consists of 
24~systems, the 18~absorbers listed in Table~\ref{tab:fits} and six DLAs from the 
literature, discussed in the last section. We have excluded the four DLAs towards 3C196, 3C336, 
0957+561A and 1345+258 (all at $z_{\rm abs} < 1.5$), for which spin temperatures are not 
quoted in the literature due to their known low covering factors [except for 3C336; \citealp{curran07a}]. 
The sample contains 14~DLAs at $z > 1.5$ and 10~at $z < 1.5$, the ``high-$z$'' and ``low-$z$''  
sub-samples, respectively (the low-$z$ DLAs are all at $z < 0.9$); the spatial extent of the core emission 
is $\lesssim 500$~pc at the DLA redshift in all cases except the $z \sim 2.039$ DLA towards 0458$-$020 
(see below).  No evidence for redshift evolution 
can be seen in Fig.~\ref{fig:f}[A], which plots the covering factor against redshift for the full sample. 
The median covering factor for the high-$z$ sub-sample is $f_{\rm med} \sim 0.685\pm 0.042$, with $f \gtrsim 0.45$ 
for all high-$z$ DLAs. The low-$z$ sample has $f \sim 0.41-1$, with a median value of $f_{\rm med} \sim 0.885 
\pm 0.094$. A Kolmogorov--Smirnov rank-1 test yields a Gaussian probability of $\sim 72$\% that 
the two sub-samples are drawn from different distributions, consistent (within $\sim 1.1\sigma$ significance) 
with the null hypothesis that they are drawn from the same distribution. Finally, the covering factors 
for the high-$z$ sample were all determined at frequencies $\lesssim \nu_{\rm 21cm}$; these are hence 
{\it lower limits}, as the spectral index of core emission is usually inverted or flat, while extended 
non-thermal emission typically has a steep spectrum, and is hence fractionally larger at lower frequencies. 
This renders the difference between the low-$z$ and high-$z$ samples even less significant.

The deconvolved sizes of the ``core'' components are $\lesssim 70$~mas for all high-$z$ DLAs in 
Table~\ref{tab:fits}, corresponding to spatial sizes of $\lesssim 500$~pc at the DLA redshift. 
Fig.~\ref{fig:f}[B] shows the spatial extent of the ``core'' emission plotted against DLA redshift; 
for the sources of Table~\ref{tab:fits}, fitted by an elliptical gaussian model, the ``size'' is 
the geometric mean of the deconvolved major and minor axes listed in Column~(9) of the table. 
In all but one of the 24~systems, the size is $\lesssim 400$~pc, which would be entirely covered 
even by small ($\gtrsim few$~kpc-sized) galaxies. The only exception is the $z \sim 2.039$ DLA towards 
0458$-$020, where \citet{briggs89} find that the entire $\sim 2''$ radio emission is covered by the 
foreground DLA, implying an absorber size $\gtrsim 17$~kpc. Note that the larger deconvolved sizes typically 
obtained at high redshifts are at least partly due to the fact that all high-$z$ DLAs were observed at 
327~MHz, where residual phase errors in the data are likely to be an issue; as mentioned earlier, the 
deconvolved sizes should be treated as upper limits.

19 of the 24~systems plotted in Fig.~\ref{fig:f} (of which ten are at $z > 1.5$) have $\ts$ 
estimates in the literature [e.g. \citealp{wolfe79,carilli96,kanekar06}; KC03]. It thus appears 
that the high $\ts$ estimates in high-$z$ DLAs are not the result of very low covering factors. 
The inferred high spin temperatures in high-$z$ DLAs could then arise due to (1)~a preponderance 
of warm \hi\ in these systems (\citealp{carilli96}; KC03), or (2)~systematically lower 
\hi\ column densities on the radio sightlines than those measured towards the optical QSO, 
due to small-scale (sub-kpc) structure in the \hi\ \citep{wolfe03b}. The spatial resolution 
of the present VLBA data is not sufficient to rule out the possibility of differences between
the optical and radio sightlines. However, good agreement (within a factor of $\sim 2$) has 
been found between \hi\ column densities measured along the same sightline from Lyman-$\alpha$ 
absorption and low-resolution \hi~21cm emission studies, both in the Galaxy \citep{dickey90} 
and in the $z \sim 0.009$ DLA towards SBS~1549+593 \citep{chengalur02}. This suggests that 
systematic large differences in \hi\ column density between radio and optical sightlines 
in DLAs are unlikely.  The high spin temperatures in high-$z$ DLAs are thus more likely 
to be the result of a larger fraction of warm HI in these absorbers.

\noindent {\bf Acknowledgments}\\
The National Radio Astronomy Observatory is operated by Associated 
Universities, Inc. under cooperative agreement with the National Science Foundation.
NK acknowledges support from the Max Planck Foundation and an NRAO Jansky Fellowship, 
and thanks Craig Walker for discussions on the VLBA analysis. Basic Research in astronomy 
at the Naval Research Laboratory is supported by the Office of Naval Research.

\bibliographystyle{mn2e}
\bibliography{ms}

\begin{thebibliography}{}

\bibitem[\protect\citeauthoryear{{Boisse}, {Le Brun}, {Bergeron} \&
  {Deharveng}}{{Boisse} et~al.}{1998}]{boisse98}
{Boisse} P.,  {Le Brun} V.,  {Bergeron} J.,    {Deharveng} J.-M.,  1998, A\&A,
  333, 841

\bibitem[\protect\citeauthoryear{{Bondi}, {Padrielli}, {Fanti}, {Ficarra},
  {Gregorini}, {Mantovani}, {Bartel}, {Romney}, {Nicolson} \& {Weiler}}{{Bondi}
  et~al.}{1996}]{bondi96}
{Bondi} M.,  {Padrielli} L.,  {Fanti} R.,  {Ficarra} A.,  {Gregorini} L.,
  {Mantovani} F.,  {Bartel} N.,  {Romney} J.~D.,  {Nicolson} G.~D.,    {Weiler}
  K.~W.,  1996, A\&A, 308, 415

\bibitem[\protect\citeauthoryear{Braun \& Walterbos}{Braun \&
  Walterbos}{1992}]{braun92}
Braun R.,  Walterbos R.,  1992, ApJ, 386, 120

\bibitem[\protect\citeauthoryear{{Bridle}, {Hough}, {Lonsdale}, {Burns} \&
  {Laing}}{{Bridle} et~al.}{1994}]{bridle94}
{Bridle} A.~H.,  {Hough} D.~H.,  {Lonsdale} C.~J.,  {Burns} J.~O.,    {Laing}
  R.~A.,  1994, AJ, 108, 766

\bibitem[\protect\citeauthoryear{{Briggs}}{{Briggs}}{1983}]{briggs83b}
{Briggs} F.~H.,  1983, AJ, 88, 239

\bibitem[\protect\citeauthoryear{Briggs \& Wolfe}{Briggs \&
  Wolfe}{1983}]{briggs83}
Briggs F.~H.,  Wolfe A.~M.,  1983, ApJ, 268, 76

\bibitem[\protect\citeauthoryear{{Briggs}, {Wolfe}, {Liszt}, {Davis} \&
  {Turner}}{{Briggs} et~al.}{1989}]{briggs89}
{Briggs} F.~H.,  {Wolfe} A.~M.,  {Liszt} H.~S.,  {Davis} M.~M.,    {Turner}
  K.~L.,  1989, ApJ, 341, 650

\bibitem[\protect\citeauthoryear{Carilli, Lane, {de Bruyn}, Braun \&
  Miley}{Carilli et~al.}{1996}]{carilli96}
Carilli C.~L.,  Lane W.~M.,  {de Bruyn} A.~G.,  Braun R.,    Miley G.~K.,
  1996, AJ, 111, 1830

\bibitem[\protect\citeauthoryear{Chengalur \& Kanekar}{Chengalur \&
  Kanekar}{2000}]{chengalur00}
Chengalur J.~N.,  Kanekar N.,  2000, MNRAS, 318, 303

\bibitem[\protect\citeauthoryear{Chengalur \& Kanekar}{Chengalur \&
  Kanekar}{2002}]{chengalur02}
Chengalur J.~N.,  Kanekar N.,  2002, A\&A, 388, 383

\bibitem[\protect\citeauthoryear{{Condon}, {Cotton}, {Greisen}, {Yin},
  {Perley}, {Taylor} \& {Broderick}}{{Condon} et~al.}{1998}]{condon98}
{Condon} J.~J.,  {Cotton} W.~D.,  {Greisen} E.~W.,  {Yin} Q.~F.,  {Perley}
  R.~A.,  {Taylor} G.~B.,    {Broderick} J.~J.,  1998, AJ, 115, 1693

\bibitem[\protect\citeauthoryear{{Curran}, {Murphy}, {Pihlstr{\"o}m}, {Webb} \&
  {Purcell}}{{Curran} et~al.}{2005}]{curran05}
{Curran} S.~J.,  {Murphy} M.~T.,  {Pihlstr{\"o}m} Y.~M.,  {Webb} J.~K.,
  {Purcell} C.~R.,  2005, MNRAS, 356, 1509

\bibitem[\protect\citeauthoryear{{Curran}, {Tzanavaris}, {Murphy}, {Webb} \&
  {Pihlstroem}}{{Curran} et~al.}{2007}]{curran07a}
{Curran} S.~J.,  {Tzanavaris} P.,  {Murphy} M.~T.,  {Webb} J.~K.,
  {Pihlstroem} Y.~M.,  2007, MNRAS, 381, L6

\bibitem[\protect\citeauthoryear{{Curran} \& {Webb}}{{Curran} \&
  {Webb}}{2006}]{curran06}
{Curran} S.~J.,  {Webb} J.~K.,  2006, MNRAS, 371, 356

\bibitem[\protect\citeauthoryear{{de Bruyn}, O'Dea \& Baum}{{de Bruyn}
  et~al.}{1996}]{bruyn96}
{de Bruyn} A.~G.,  O'Dea C.~P.,    Baum S.~A.,  1996, A\&A, 305, 450

\bibitem[\protect\citeauthoryear{{Dickey} \& {Lockman}}{{Dickey} \&
  {Lockman}}{1990}]{dickey90}
{Dickey} J.~M.,  {Lockman} F.~J.,  1990, ARA\&A, 28, 215

\bibitem[\protect\citeauthoryear{{Douglas}, {Bash}, {Bozyan}, {Torrence} \&
  {Wolfe}}{{Douglas} et~al.}{1996}]{douglas96}
{Douglas} J.~N.,  {Bash} F.~N.,  {Bozyan} F.~A.,  {Torrence} G.~W.,    {Wolfe}
  C.,  1996, AJ, 111, 1945

\bibitem[\protect\citeauthoryear{{Frey}, {Gurvits}, {Altschuler}, {Davis},
  {Perillat}, {Salter}, {Aller}, {Aller} \& {Hirabayashi}}{{Frey}
  et~al.}{2000}]{frey00}
{Frey} S.,  {Gurvits} L.~I.,  {Altschuler} D.~R.,  {Davis} M.~M.,  {Perillat}
  P.,  {Salter} C.~J.,  {Aller} H.~D.,  {Aller} M.~F.,    {Hirabayashi} H.,
  2000, PASJ, 52, 975

\bibitem[\protect\citeauthoryear{Kanekar \& Chengalur}{Kanekar \&
  Chengalur}{2001}]{kanekar01a}
Kanekar N.,  Chengalur J.~N.,  2001, A\&A, 369, 42

\bibitem[\protect\citeauthoryear{Kanekar \& Chengalur}{Kanekar \&
  Chengalur}{2003}]{kanekar03}
Kanekar N.,  Chengalur J.~N.,  2003, A\&A, 399, 857

\bibitem[\protect\citeauthoryear{Kanekar, Chengalur \& Lane}{Kanekar
  et~al.}{2007}]{kanekar07}
Kanekar N.,  Chengalur J.~N.,    Lane W.~M.,  2007, MNRAS, 375, 1528

\bibitem[\protect\citeauthoryear{Kanekar, Subrahmanyan, Ellison, Lane \&
  Chengalur}{Kanekar et~al.}{2006}]{kanekar06}
Kanekar N.,  Subrahmanyan R.,  Ellison S.~L.,  Lane W.~M.,    Chengalur J.~N.,
  2006, MNRAS, 370, L46

\bibitem[\protect\citeauthoryear{Lane, Briggs \& Smette}{Lane
  et~al.}{2000}]{lane00}
Lane W.~M.,  Briggs F.~H.,    Smette A.,  2000, ApJ, 532, 146

\bibitem[\protect\citeauthoryear{{Rengelink}, {Tang}, {de Bruyn}, {Miley},
  {Bremer}, {R{\" o}ttgering} \& {Bremer}}{{Rengelink}
  et~al.}{1997}]{rengelink97}
{Rengelink} R.~B.,  {Tang} Y.,  {de Bruyn} A.~G.,  {Miley} G.~K.,  {Bremer}
  M.~N.,  {R{\" o}ttgering} H.~J.~A.,    {Bremer} M.~A.~R.,  1997, A\&AS, 124,
  259

\bibitem[\protect\citeauthoryear{{Wilkinson}, {Readhead}, {Anderson} \&
  {Purcell}}{{Wilkinson} et~al.}{1979}]{wilkinson79}
{Wilkinson} P.~N.,  {Readhead} A.~C.~S.,  {Anderson} B.,    {Purcell} G.~H.,
  1979, ApJ, 232, 365

\bibitem[\protect\citeauthoryear{{Wolfe}, {Briggs} \& {Jauncey}}{{Wolfe}
  et~al.}{1981}]{wolfe81}
{Wolfe} A.~M.,  {Briggs} F.~H.,    {Jauncey} D.~L.,  1981, ApJ, 248, 460

\bibitem[\protect\citeauthoryear{{Wolfe}, {Broderick}, {Johnston} \&
  {Condon}}{{Wolfe} et~al.}{1978}]{wolfe78}
{Wolfe} A.~M.,  {Broderick} J.~J.,  {Johnston} K.~J.,    {Condon} J.~J.,  1978,
  ApJ, 222, 752

\bibitem[\protect\citeauthoryear{Wolfe \& Davis}{Wolfe \&
  Davis}{1979}]{wolfe79}
Wolfe A.~M.,  Davis M.~M.,  1979, AJ, 84, 699

\bibitem[\protect\citeauthoryear{Wolfe, Gawiser \& Prochaska}{Wolfe
  et~al.}{2003}]{wolfe03b}
Wolfe A.~M.,  Gawiser E.,    Prochaska J.~X.,  2003, ApJ, 593, 235

\bibitem[\protect\citeauthoryear{Wolfe, Gawiser \& Prochaska}{Wolfe
  et~al.}{2005}]{wolfe05}
Wolfe A.~M.,  Gawiser E.,    Prochaska J.~X.,  2005, ARA\&A, 43, 861

\bibitem[\protect\citeauthoryear{Wolfe, Turnshek, Smith \& Cohen}{Wolfe
  et~al.}{1986}]{wolfe86}
Wolfe A.~M.,  Turnshek D.~A.,  Smith H.~E.,    Cohen R.~D.,  1986, ApJS, 61,
  249

\bibitem[\protect\citeauthoryear{{York}, {Kanekar}, {Ellison} \&
  {Pettini}}{{York} et~al.}{2007}]{york07}
{York} B.~A.,  {Kanekar} N.,  {Ellison} S.~L.,    {Pettini} M.,  2007, MNRAS,
  382, L53

\end{thebibliography}

\end{document}